\documentstyle[aps,twocolumn,psfig]{revtex}
\begin{document}
\draft

\title{The transition from a classical to a quantum world as a
passage from extensive to non-extensive thermodynamics} \author{Luigi
Palatella$^{1}$ and Paolo Grigolini$^{1,2,3}$}
\address{$^{1}$Dipartimento di Fisica dell'Universit\`{a} di Pisa,
Piazza
Torricelli 2, 56127 Pisa, Italy }
\address{$^{2}$Center for Nonlinear Science, University of North
Texas, P.O. Box 305370, Denton, Texas 76203 }
\address{$^{3}$Istituto di Biofisica del Consiglio Nazionale delle
Ricerche, Via San Lorenzo 26, 56127 Pisa, Italy } \date{\today}
\maketitle

\begin{abstract}
We study the thermodynamical properties of the quantum kicked
rotator, coarsened by an external fluctuation with a weak intensity
$D$, by means
of the Tsallis entropy
with a
changing entropic index $q$.
The genuine
entropic index, corresponding
to given values of $D$ and $\hbar$ is that making the Tsallis entropy
 increase linearly in time, and it is proved to
become $q <1 $ for suitably large values of $\hbar $: This indicates
a subdiffusional regime which,
in turn, signals
the occurrence of quantum localization. Thus the process of Anderson
localization is shown to be compatible with a thermodynamical
representation provided that a non-extensive form of entropy is used.
\end{abstract}


The study of the connection between dynamics and thermodynamics is
making significant progresses along the lines of the seminal work of
Krylov~\cite{Krylov}. Interesting attempts are currently being made
at relating
the Kolmogorov-Sinai (KS) entropy~\cite{KS} to the thermodynamical
entropy. Of remarkable interest are the work of Gaspard, relating the
KS entropy for a dilute gas to the standard thermodynamical entropy
per unit volume of an ideal gas~\cite{Gaspard}, and the more recent
paper by Dzugutov, Aurell and Vulpiani~\cite{DAV98}, who express the
KS
entropy of a simple liquid in terms of the excess entropy, namely,
the difference between the thermodynamical entropy and that of the
ideal gas at the same thermodynamical state.

On the other hand, the realization of Boltzmann's
dream~\cite{Lebowitz} implies the Gibbs entropy

\begin{equation}
S_{G}(t) = -\int \mbox{\rm d\boldmath$X$}
\rho(\mbox{\boldmath$X$},t)\log(\rho(\mbox{\boldmath$X$},t)) \ ,
\label{gibbs}
\end{equation}
corresponding to a given nonequilibrium initial condition, to quickly
increase towards the equilibrium condition established by Boltzmann's
entropy. The time evolution of the Liouville density $\rho(t)$ is, on the
contrary,
unitary, thereby making the Gibbs entropy time independent.
As a consequence, in accordance with the view of Krylov~\cite{Krylov}
a coarsening process is required. Since, the second principle
~\cite{huang} requires the system to be thermally isolated from the
environment, and not necessarily totally
isolated,
we deem to
be acceptable to
perturb its quantum equation with a stochastic force of
weak intensity $D$~\cite{ZP94}:
With heuristic arguments these authors ~\cite{ZP94} prove that this
weak perturbation makes the Gibbs
entropy
of Eq.~(\ref{gibbs}) reach the maximum value established by
Boltzmann's
entropy in the short time scale $1/\lambda$, with $\lambda$ denoting
the Lyapunov coefficient. This heuristic prediction has been more
recently checked numerically by Pattanayak and Brumer~\cite{PB97} who
used rather than the conventional Gibbs entropy another form of
\emph{extensive} entropy, the R\'{e}nyi entropy~\cite{BS93}. Thus,
the entropy increase has a rate fixed by $\lambda_{2}$, where
$\lambda_{2}$ is a generalized Lyapunov coefficient coinciding with
$\lambda$ in the ideal case of strong
chaos.

The important problem of how to deal with the case of \emph{non
extensive} thermodynamics
has been addressed about ten years ago by
Tsallis~\cite{CONSTANTINO88}, with a generalized
entropy that, expressed in terms of the Liouville density
$\rho(t)$, reads: \begin{equation}
S_{q}(t)
= \frac{1-\int\rm d \mbox{\boldmath$X$}
\rho(\mbox{\boldmath$X$},t)^{q}}{q-1} .
\label{tsallis}
\end{equation}
As an important benefit this entropy makes it possible to
extend the KS method of numerical analysis to the fractal
dynamics\cite{TPZ97}.The authors of Ref.\cite{TPZ97} generalized the
KS entropy by replacing the conventional Shannon expression, formally
equivalent to the structure proposed by Gibbs, with the non-extensive
form of Tsallis. In the special case
$q=1$, when Eq.(\ref{tsallis}) yields Eq.(\ref{gibbs}), the generalized
 KS entropy, denoted by $K_{q}$ , becomes identical to the conventional
KS entropy\cite{Gaspard,DAV98}.

We establish a connection between $K_{q}$ and the generalized entropy
of Eq.~(\ref{tsallis}) with the same coarsening procedure as that
originally proposed
by Zurek and Paz~\cite{ZP94}. However, to adapt it to the special
case of fractal dynamics
we found to be convenient
to use the intuitive picture
by Zaslavsky ~\cite{Zaslavsky94}. We adopt an initial
condition compatible with classical physics so as to
generate a trajectory
that will explore in due time a given region of the phase space
according to the prescriptions of
classical mechanics. Let us call $R(t)$ the portion of this region
not yet visited at the given time $t$ by the classical trajectory. It
is evident that in general this quantity will be a decreasing function
of time. For generality, on the basis of the arguments used in
Ref.\cite{TPZ97} to generalize the KS entropy, we express the general
form of this time evolution as:
\begin{equation}
R(t)
= \frac{R(0)}{[1 + \lambda_{q}(1 - q)t]^{\frac{1}{(1 - q)}}} ,
\label{tsalliszaslavsky}
\end{equation}
where $\lambda_{q}$ denotes a generalized Lyapunov coefficient.
Let us imagine that at the very same moment when the initial
conditions
are established a weak fluctuation process of intensity $D$ is
switched on. In the absence of fluctuations a set of trajectories,
with the same initial condition, will behave as a single trajectory,
even if this single trajectory will be made to look erratic-like by
deterministic chaos. The action of an even extremely weak
fluctuation, in the presence of chaos, will force the trajectories of
this set to spread over the proper phase-space region even if
all of them are assumed to be given {\em exactly } the same initial
condition.
This weak stochastic force serves the purpose of defining in a
precise way the time at which the exploration of the phase space can
be regarded as being completed. We call this time $t_{CG}$ and we
define it
by means of the following relation:

\begin{equation}
R(t_{CG})=Dt_{CG}. \label{fundamentalequation}
\end{equation}
We note that the solution of this fundamental equation depends on the
nature of the dynamical process under study. In the case of strong
chaos\cite{TPZ97} we have $q=1$ and, consequently,

\begin{equation}
t_{CG}\approx \frac{2}{\lambda _{1}}\log(1/\sqrt{D}), \label{ordinarytcg}
\end{equation}
which coincides with the estimate made by Zurek and Paz to determine
the onset time of thermodynamics. However, in the
case of highly correlated dynamical processes, with $q<1$ as in the
case here under study, we obtain from Eq.~(\ref{fundamentalequation})

\begin{equation}
t_{CG}\approx \frac{1}{^{D^{\theta }}}, \label{elenarenato}
\end{equation}
where

\begin{equation}
\theta = (1-q)/(2-q). \label{thetruth}
\end{equation}

This result gives
the misleading impression that the thermodynamical regime is
dramatically
postponed in the case of dynamical processes with no time scale. It
is important to stress that this
postponment only concerns the onset of ordinary thermodynamics. In
fact, a weak stochastic force
perturbing the dynamics with long-range correlations responsible for
anomalous diffusion has the key effect, as shown in Ref.\cite{FMG96},
of producing a crossover from anomalous to ordinary diffusion at the
time scale of Eq.(\ref{elenarenato}). However, according to the
perspective established by Tsallis\cite{CONSTANTINO88}, also the
regime of anomalous diffusion is given a thermodynamical significance
if the proper entropic index
$q\neq1$ is used. Therefore we expect that
the coarsening stochastic force
producing the crossover from anomalous to ordinary
diffusion at the time scale of Eq.(\ref{elenarenato}) results
in the
fast increase of the entropy $S_{q}$
of Eq.(\ref{tsallis}). This means that a transition
from the regime of Eq. (\ref{elenarenato})
to that of Eq. (\ref{ordinarytcg}) (with $q\neq1$) occurs and
consequently that the entropy increases with
the rate
$1/\lambda_{q}$.

In this letter we support this conjecture in a case where the origin
of long-range correlation is quantum. As pointed out by Zurek and
Paz\cite{ZP94}, the time scale of Eq.(\ref{ordinarytcg}) has to be
compared to

\begin{equation}
t_{Q}\approx \frac{1}{\lambda}\log(\frac{1}{\hbar}), \label{tq}
\end{equation}
which corresponds to the onset of long-range quantum correlations. The
transition from quantum to classical physics takes place when $\sqrt{D}$ is
made larger than $\hbar$. Rather than focusing our attention on
values of $\hbar$ so small as to ensure the classicality
condition\cite{PB97}, we explore the transition region, and the
long-range correlations of the region $\hbar >\sqrt{D}$.The case study is
given by the quantum
kicked rotor. This means the Hamiltonian

\begin{equation}
H(t)= - \frac{\hbar ^{2}}{2I}\frac{\partial ^{2}}{\partial \theta
^{2}} +\epsilon \cos \theta \sum_{n=-\infty }^{\infty }\delta (t-nT),
\label{kickedrotorhamiltonian}
\end{equation}
where $I$ is the rotor moment of inertia and $\theta $ is the rotor
angular coordinate. This Hamiltonian results in the quantum map

\begin{equation}
\langle \theta |\phi (n+1)\rangle =\exp (i\frac{\hbar T
}{2I}\frac{\partial ^{2}}{\partial \theta ^{2}})\exp (-ik\cos \theta
) \langle \theta|\phi (n)\rangle , \label{quantummap} \end{equation}
implying the classical control parameter $K$\cite{FELIX90} to be
expressed as: $K = \frac{\hbar T}{2I} k$.

Here we adopt two different numerical methods to evaluate the entropy
time evolution. The former is the same as that designed by Pattanayak
and Brumer, based on a Fourier transform technique\cite{PB97}. The
latter, henceforth referred to as a statistical ensemble (SE) method,
is based on the study of the time evolution of $N$ independent
systems, all of them driven by Eq.(\ref {quantummap}), whith a random
ingredient artificially added by replacing
$exp \left (-i \frac{p^{2}}{2I\hbar} \right )$ with
$exp \left (-i \frac{(p+F(n))^{2}}{2I\hbar}\right )$, where $F(n)$ is
a computer generated random process. This computed generated random
process is introduced to play the same coarsening role as the earlier
mentioned stochastic force of the
picture of Zurek and Paz\cite{ZP94}.
For this reason the variance of
this random process $F(n)$ is called
$D$, is assigned very small values and is referred to throughout as
intensity of the stochastic force.
The statistical density matrix is then evaluated by means of the
ordinary statistical average:

\begin{equation}
\rho (n)=\frac{1}{N}\sum_{i}^{N}|\phi _{i}(n)\rangle \langle \phi
_{i}(n)|, \label{statisticalaverage} \end{equation}
usually setting  $N = 10$.
The same method has been used
in an earlier paper\cite{BGLR96}
and proved to produce results equivalent to those of the
literature on
this subject. It is important to
stress that the ingenious
Fourier method of
Ref.\cite{PB97} establishes a connection with the system dynamics
unperturbed by the stochastic force, much in the spirit of the
linear response theories. The method here adopted, on the contrary,
allows us to go beyond the second-order approximation, and makes it
possible to detect, as we shall see, the saturation effects
determined by the steady action of the stochastic force.
\begin{figure}
\centerline{
\vbox{
\psfig{figure=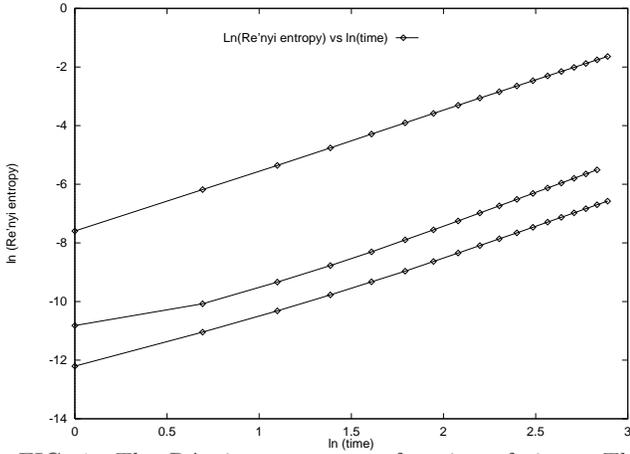,width= 3.4 in} 
}}
\caption[]{The R\'{e}nyi entropy as a function of time. The values
of $\hbar$ adopted are from the the top to the bottom curve:
$\hbar=0.01, 0,05, 0.1$. We also set $K = 7.1$ }
\label{Fig1}
\end{figure}

In Fig.\ref{Fig1} we show the increase of R\'{e}nyi entropy as a
function of time
for different values of $\hbar $. This calculation was done using the
same method as that of Ref.\cite{PB97}
with a choice of parameters
corresponding to the
quantum condition that we plan to explore by using the SE method. We checked
that the SE method yields tha same results as those illustrated in
 Fig.\ref{Fig1}, if the same parameters are used and we set $\sqrt{D}=0.002$.
We note that Pattanayak and Brumer\cite{PB97} considered also values
of $\hbar$ much smaller than the value $\hbar = 0.01$ here
considered. This is the reason why we depart from the regime of
exponential increase, searched and found by those authors, and we
rather get
the power law increase: $S(t)=cons\ t^{\alpha }$ with $\alpha \approx
2.5$. This property is compatible with the numerical results obtained
by Pattanayak
and Brumer at their large values of $\hbar$. These authors, however,
seemingly did not consider this behavior to be thermodynamically
relevant.
On the basis of our earlier remark, on
the contrary, we are here in a position to disclose the
thermodynamical nature of this regime, in spite of the fact that one
might judge
it dominated by the quantum mechanical coherence, and by the Anderson
quantum
localization\cite{PB97} as well.

\begin{figure}
\centerline{
\vbox{
\psfig{figure=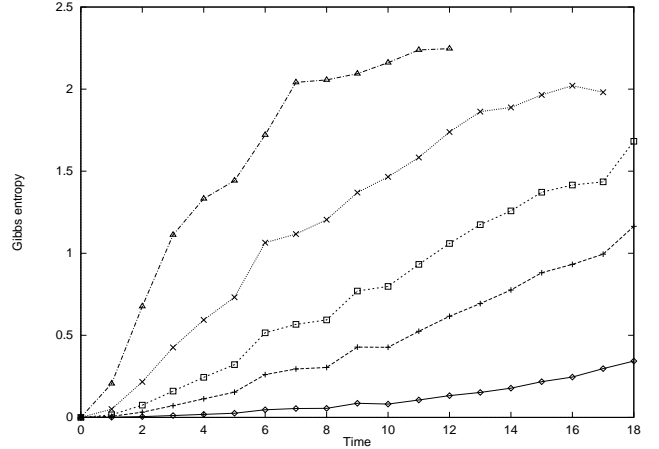,width= 3.4 in} 
}}
\caption[]{The Gibbs entropy as a
function of time for $\hbar= 0.01$, $K =7.1$ and different values
of the noise intensity $D$. The
values of $D$ changes from the bottom to the top curve as follows:
$\sqrt{D} = 0.002, 0.006, 0.01, 0.02, 0.05$.} \label{Fig2}
\end{figure}

\begin{figure}
\centerline{
\vbox{
\psfig{figure=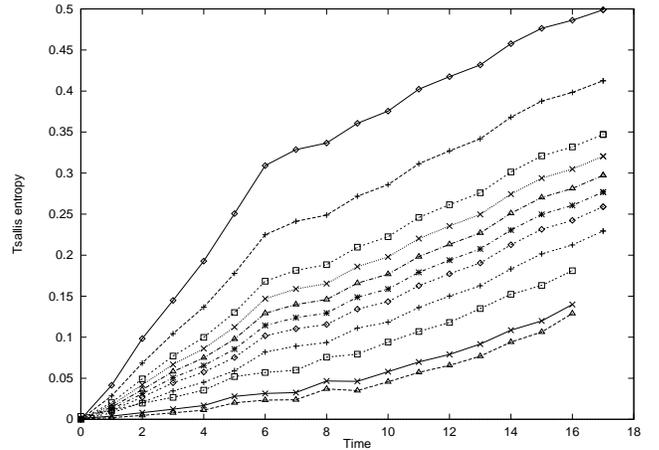,width= 3.4 in} }}
\caption[]{The Tsallis entropy as a
function of time. The values of the parameters used are
$\hbar=0.01$, $K = 7.1$ and $\sqrt{D}=0.002$.
The entropic
index $q$ changes from the top to the bottom curve as follows: $q =
0.155, 0.205, 0.255,$ $ 0.280, 0.305, 0.330, 0.335, 0.405,$
$ 0.530, 0.555$
and $0.805$.}
\label{Fig3}
\end{figure}

We are convinced that at $\hbar = 0.01$ we are in the presence of a
condition
equivalent to that of fractal dynamics with no time scale, and thus
corresponding to an entropic index $q\neq1$. However, with increasing
the intensity of the stochastic force we can provoke a transition
from quantum to classical physics which, as earlier pointed out, must
take place at $\sqrt{D} = \hbar$. Fig.\ref{Fig2} shows the following
very remarkable property. At small values of $D$ the curve $S_{G}(t)$
is
characterized by a positive second-order derivative with respect to
time (convex curve). With increasing the noise intensity a transition
from concavity to convexity is produced at about the value of $D$
($\sqrt{D}\approx0.01$) corresponding to the passage from quantum to
classical physics.
We assume that the transition from convexity to concavity has a
thermodynamical relevance, and this leads us to find still
more interesting results by
plotting the Tsallis entropy as a function of time for different
values of the entropic index $q$. This is shown in Fig.\ref{Fig3},
which illustrates the remarkable property that decreasing $q$ has
effects similar to those obtained by increasing $D$ (see
Fig.\ref{Fig2}). The critical value of $q$ corresponding to this
transition is $q\approx0.33$.


\begin{figure}
\centerline{
\vbox{
\psfig{figure=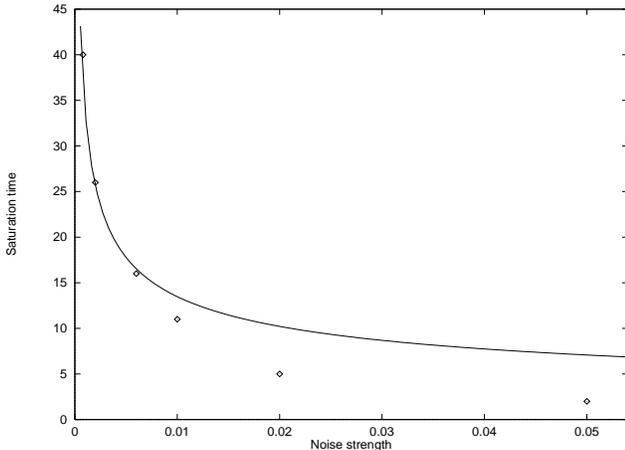,width= 3.4 in}
 }}
\caption[]{The saturation time $t_{S}$ as
a function of the noise strength $\sqrt{D}$.
 The full line corresponds to
the prediction
 of Eqs.~(\ref{elenarenato}) and~(\ref{thetruth}) with the critical
entropic index $q$ determined by the numerical results
illustrated in Fig.\ref{Fig3}.
 }
\label{Fig4}
\end{figure}

To make our finding still more impressive we plot in Fig.\ref{Fig4} the
saturation time $t_{S}$ as a function of the noise intensity $D$. To
derive the results of Fig.\ref{Fig4} we defined the saturation time
as the time at which the curve $S_{G}(t)$ reaches the 50\%\ of its
maximum value. The remarkable fact illustrated by this figure is that
the full line corresponds to the theoretical prediction of
Eq.(\ref{elenarenato}) with the index $\theta$ given by
Eq.(\ref{thetruth}) and the index $q$ determined by the critical
value established by Fig.\ref{Fig3}. As we have seen, this means
$q\approx0.33$, and, through Eq.(\ref{thetruth}), $\theta\approx0.4$,
which is in fact the power index of the full curve of Fig.\ref{Fig4}.
Note that the choice made to derive Eq. (\ref{thetruth}) is to some
extent arbitrary. However, the key index $\theta$ can also be derived
from $\theta = 1/\alpha$, where $\alpha$ refers to the result of
Fig.\ref{Fig1}, $S(t)=cons\ t^{\alpha }$ with $\alpha
\approx 2.5$, coinciding with the earlier prediction. We also note
that
after the transition to the classical regime the decrease of $t_{S}$
with the increase of $D$ becomes faster, as made necessary by the
passage from the regime of Eq.(\ref{elenarenato}) to that of
Eq.(\ref{ordinarytcg}).

On the basis of the earlier arguments it is not a surprise that a
condition compatible with the generalized expression of the KS
entropy is found at $q\neq 1$. One might wonder, however, why
$q < 1$ . This can be easily explained by adopting the entropic
arguments of Tsallis illustrated
in Refs.\cite{tsallisphysica} and \cite{tsallisbukman96}.
 In \cite{tsallisbukman96} the subdiffusional
character corresponding to the entropic index $q$ is proved by means
of a nonlinear Fokker-Planck equation. However, there are good reasons
to believe that this is a quite general property of the entropic
index
$q$\cite{condmatt} (see also\cite{note}). Thus we conclude that the
fundamental results illustrated by Fig.\ref{Fig3} and Fig.\ref{Fig4}
 signal a sort of quantum-mechanically
induced subdiffusion. This conclusion agrees very well with the well
known fact that the quantum mechanical kicked rotor is characterized
by the phenomenon of Anderson localization\cite{FELIX90}. It is well
known that the localization process has a rate proportional to about
$\hbar^{2}$. As a consequence we expect that the entropic index $q$
as a function of $\hbar$ tends to $1$ with decreasing $\hbar$,
in agreement with
the numerical observation that the values $\hbar = 0.10,
0.05$ and $0.01$ yield $q = 0.28, 0.30$ and $0.33$, respectively.
We thus reach the interesting conclusion
that even the process of Anderson localization is compatible with a
thermodynamical treatment, provided that a proper thermodynamical
indicator is adopted: This is the
Tsallis entropy. Furthermore the adoption of the SE method of
calculation makes it possible to show that if the Liouville density
of Eq.(\ref{tsallis}) is coarsened by means of a weak stochastic
force and the proper choice of the entropic index $q$ is made, then
the time derivative of $S_{q}(t)$ becomes constant, as prescribed by
the generalization of the KS entropy proposed by the authors of
Ref.\cite{TPZ97}.


\begin{references}
\bibitem{note} In Ref.~\cite{TPZ97} it was shown that the adoption of
the Tsallis entropy yields a natural generalization of the KS
entropy, and this, in turns yields the
important result:\( \lim_{\Delta x(0) \rightarrow 0} \frac{\Delta
x(t)}{\Delta x(0)}
= [1 + \lambda_{q}t(1-q)]^{\frac{1}{1-q}} \). Note that this
prescription for $q>1$ implies a divergence at finite time, and,
consequently a departure
of initially very close trajectories from one another faster that
an exponential departure. The case $q<1$ yields a power law
dependence on time, which can be interpreted as slower than the
exponential departure. It is well known that the case $q=1$ results
in ordinary Brownian diffusion. In Ref.~\cite{condmatt} it was shown
that the case $q>1$ leads to superdiffusion. Consequently, the case
$q<1$, with a departure of the trajectories slower than exponential,
ought to result, in
general, in subdiffusion, regardless whether the nonlinear diffusion
equation of Ref.~\cite{tsallisbukman96} applies or not.
\end{references}
\end{document}